 \documentclass[final,3p,times]{elsarticle}
\usepackage{amssymb}
\usepackage{amsmath}
\usepackage{xcolor}
\journal{Chaos, Solitons and Fractals}

\begin{document}

\begin{frontmatter}



\title{Kink-Driven Chimera Motion with Quantized Velocity in a Chain of Interacting Particles}


\author[inst1]{Maxim~I.~Bolotov\corref{cor1}}
\ead{maksim.bolotov@itmm.unn.ru}
\cortext[cor1]{Corresponding author}

\affiliation[inst1]{organization={Department of Control Theory, Lobachevsky State University of Nizhny Novgorod},
            addressline={Gagarin~Ave.~23}, 
            city={Nizhny Novgorod},
            postcode={603022}, 
            country={Russia}}

\affiliation[inst2]{organization={Gaponov-Grekhov Institute of Applied Physics of the Russian Academy of Sciences},
            addressline={Ul’yanova Str.~46}, 
            city={Nizhny Novgorod},
            postcode={603950}, 
            country={Russia}}

\author[inst1]{Lev~A.~Smirnov}
\author[inst2,inst1]{Vasily~A.~Kostin}
\author[inst1]{Grigory~V.~Osipov}

\begin{abstract}
We investigate chimera synchronization of internal oscillator states in a ring of interacting particles, using the damped dc-driven Frenkel--Kontorova chain model as an example. In a system with a spatially periodic potential, a dc external force, and dissipation, kinks spontaneously emerge and stabilize. We show that these kinks induce and govern a collective motion of the entire chimera pattern of internal states along the ring. In particular, the average velocity of this motion depends linearly on the number of kink pairs. This number is effectively determined by localized initial perturbations of particle positions, thereby opening a pathway for controlling macroscopic transport through microscopic excitations.
\end{abstract}




\end{frontmatter}



%---------------------------------------------------------------------------%
\section{Introduction}
\label{sec:Introduction}
The dynamics of active (self-propelling) particles constitute a core area of modern statistical physics focusing on non-equilibrium processes and spanning the fields of physics, chemistry, biology, and materials science~\cite{Reimann2002,Hanggi2009,Reichhardt2017}. Models of active particle provide a universal framework for modeling fundamental phenomena such as self-organization, collective behavior, and phase transitions under conditions of strong fluctuations and dissipation~\cite{Bechinger2016}. Traditionally, the dynamics of active particles have been explored within stochastic models in which noise plays a defining role in inducing particle motion~\cite{Vicsek2012}. These models successfully describe a wide spectrum of phenomena, ranging from microscopic active Brownian particles~\cite{lowen2020} to macroscopic analogs such as collective animal motion~\cite{Vicsek2012}, traffic flows~\cite{kerner2019}, and pedestrian dynamics~\cite{helbing2013}. However, in the high-activity limit, where particle velocities can be considered nearly constant, a deterministic description of dynamics becomes both practical and appropriate~\cite{aranson2022}.

%In contrast to their extensively studied stochastic counterparts, deterministic regimes of active particles remain significantly less explored.
Most early works on such deterministic descriptions focused on periodic attractors~\cite{ebeling1999, kruk2018}. Recently, however, interest in this topic has been revived, spurred by investigations of collective motion in disordered media~\cite{peruani2018}. Systems that combine particle motion and internal oscillator dynamics---so-called swarmalators~\cite{keeffe2017}---are of particular interest in this context. These hybrid objects exhibit a rich spectrum of phenomena, including collective synchronization, pattern formation, and self-assembly~\cite{snezhko2011, yan2012,keeffe2019}, and present wide opportunities for applications such as targeted drug delivery~\cite{Ebbens2016,Dietrich2018,Fehlinger2023,Cereceda2024} and high-precision particle sorting~\cite{Agrawal2020}.

In analysis of coupled oscillators, including mobile ones, current attention is focused on complex synchronization regimes such as cluster states and, in particular, chimera states---regimes characterized by coexistence of synchronized and desynchronized subsystems within ensembles of identical elements~\cite{Kuramoto2002,Abrams2004}. {The chimera states were found in systems of globally coupled oscillators and in systems with nonlocal interactions~\cite{Kuramoto2002,Abrams2004,smirnov2017,omel2018, parastesh2021}. First, the coexisting synchronous and asynchronous dynamics was reported for continuous medium of identical oscillators with nonlocal interactions~\cite{Kuramoto2002}. The coexistence of coherent and incoherent populations
had been studied even earlier in nonlocally or globally
coupled oscillators by Kuramoto~\cite{kuramoto1995,kuramoto1997} and others~\cite{kaneko1990, Nakagawa1993, Chabanol1997, Nakao1999} in different setups. Reviews~\cite{omel2018,parastesh2021} detail the systems supporting chimera states and their various classifications, but precisely defining regimes remains a challenging tasks~\cite{Haugland2021,Ferre2023}.}

{
For symmetric kernels of nonlocal coupling, the synchronous and asynchronous clusters are localized in stationary regions. However, this may change for asymmetric kernels, for which the chimera states acquire drift~\cite{Xie2014, Omelchenko2019, Omelchenko2023, Smirnov2024}, as well as other traveling and drifting structures may arise~\cite{Smirnov2024}.}

The motion of active particles plays a dual role: in some cases, it acts as an effective source of noise, disrupting chimera states and accelerating global synchronization~\cite{wang2019}, while in others, it acts to stabilize inhomogeneous structures, including chimera states~\cite{smirnov2021}. A similar duality is observed in systems with time-delayed coupling~\cite{petrungaro2019}.

In this work, we investigate the nontrivial role of particle mobility and multiple time scales in the formation of complex collective regimes in a system of deterministic active particles with an internal oscillatory degree of freedom. To this end, we consider an ensemble of Newtonian particles moving in a periodic substrate potential under the influence of a constant force and dissipation, and possessing internal phase dynamics. In our previous work~\cite{Bolotov2025}, we showed that a regime of selective drift motion emerges in such a system, wherein the phase dynamics of the ensemble self-organizes into a chimera state. This is accompanied by a hierarchical chimera: the coexistence of two chimera subsystems within the groups of flying-through (transit) and stationary particles.

Within this model, we uncover and examine an alternative dynamical scenario results from nearest-neighbor interactions. We identify a novel directed transport mechanism, driven by the emergence and propagation of kinks~\cite{braun2004, braun1998, floria1996, tekic2016} in the system of interacting particles. Unlike the previously studied running motion of non-interacting particles that leads to chimera formation, this scenario reveals a fundamentally distinct transport mechanism associated with the collective dynamics of particle displacements. The key feature of this scenario is the quantization of the chimera drift velocity, which takes on discrete values determined by the initial particle states that define the system's topological excitation. This established link between topological excitations and chimera synchronization, as well as the demonstrated linear dependence of transport characteristics on the number of kinks, broadens the understanding of collective transport mechanisms in periodic potentials and opens new avenues for controlling active matter motion.
%---------------------------------------------------------------------------%
\section{Model}~\label{sec:model}

We consider the damped dc-driven Frenkel--Kontorova (FK) model coupled through particle positions to the Kuramoto--Battogtokh (KB) system,
\begin{gather}
\mu \ddot{x}_n + \lambda \dot{x}_n + \psi \sin\left( \frac{2\pi N x_n}L \right) = \gamma + \chi (x_{n-1} - 2x_n + x_{n+1}),
\label{eq:main_x}\\
\dot{\varphi}_n = \sum_{n'=1}^{N} G\big(x_{n'} - x_n\big) \sin\!\left(\varphi_{n'} - \varphi_n - \alpha\right).
\label{eq:main_phi}
\end{gather}
Here, $x_n(t)$ and $\varphi_n(t)$ are the position and the internal phase of the $n$th particle, respectively, $n = 1, 2, \dots, N$, $N$ is the total number of identical particles constituting the system, $\mu$ is the particle mass, $\lambda > 0$ is the friction coefficient, $\psi$ is the amplitude of the spatially periodic force component, $\gamma$ is the uniform dc external force, $\chi > 0$ is the elastic (coupling) constant, $\alpha$ is the phase shift, and $G(x)$ is the kernel defining nonlocal phase interaction.
In what follows, we use the wrapped version
\begin{equation}
G(x) = \frac{\kappa \cosh\!\left( \kappa L \left|\{x/L\}-1/2\right| \right)}{2 \sinh(\kappa L/2)},
\label{eq:ker}
\end{equation}
of the exponential kernel~\cite{smirnov2017}, which can describe, for example, chemotaxis in living systems~\cite{tanaka2007}.
Here, the figure brackets $\{z\} = z - \lfloor z \rfloor$ denote the fractional part, and the parameter $\kappa$ sets the spatial scale of the phase interaction.

{Thus, in In our framework, the damped dc-driven FK model is independent of the KB system dynamics. That is, the coupling is strictly unidirectional: the FK particle positions provide input to the KB network without receiving any feedback from it.}
Equation~\eqref{eq:main_x} governs the particle motion, while~\eqref{eq:main_phi} describes the oscillatory dynamics of the internal particles states influenced by particle motion through interaction kernel depending on the relative particle positions.
Without internal phase dynamics equation~\eqref{eq:main_x} finds applications in the theory of dislocations and adsorption in crystals~\cite{braun1998, Atkinson1965}, the dynamics of domain walls in magnets~\cite{Cowley1976, belim2021}, modeling of biopolymers~\cite{yakushevich2006}, as well as for describing collective motion in active media~\cite{miyake1996, tanaka2007} and particles in optical traps~\cite{abbott2019}.

{The two following sections describe the independent dynamics of the separate subsystems governing particle motion and their phases.}

\subsection{{Particle motion dynamics}}
We consider particle moving on a ring with $x_n \in [0, L)$ and periodic boundary conditions. In the unperturbed state particles are stationed equidistantly with each particle sitting at the bottom of its own potential well, which is a classical case for the FK model. We also consider periodic particle ordering for nearest-neighbor interaction term with $x_0 \equiv x_N$ and $x_{N + 1} \equiv x_1$. Note that here nearest neighbor is meant in terms of particle ordering, not the real distance on the ring.
It is convenient to use the scaled elastic constant $\eta = \chi L^2/\mu N^2$ which corresponds to the squared maximal group velocity of small (linear) perturbations in the FK system.
By rescaling spatial coordinate, one can always set $L = 1$ without loss of generality, as we do below.

Equation~\eqref{eq:main_x} supports a variety of spatial regimes: in addition to the collective transit motion of particles, the system admits structured solutions in the form of propagating kinks~\cite{tekic2016, fitzgerald2016}. In lattices, these solutions correspond to localized defects that move along the chain. This type of motion is characterized by velocities an order of magnitude lower than those of transit motion and, as we show below, leads to new effects in the dynamics of the coupled phase variables. Below we identify the features of chimera state motion in the phase dynamics~\eqref{eq:main_phi} when kinks are present in the system~\eqref{eq:main_x}.

\subsection{{Phase dynamics for stationary and mobile particles}}
For fixed particle positions at the minima of the substrate potential $U(x) = -(\psi L/2\pi N) \cos(2\pi N x/L) - \gamma x$, system~\eqref{eq:main_phi} exhibits two main stable states: complete synchronization and {stationary} chimera states---regimes of coexistence of synchronized and unsynchronized clusters~\cite{Abrams2004, Wolfrum2011}. {These modes are statistically observable under random initial conditions, when the initial phases are distributed uniformly over $[-\pi, \pi]$.} The lifetime of a chimera state grows exponentially with $N$, making it metastable for large ensembles. {After transient process, the synchronous cluster of the chimera mode remains localized and does not exhibit drift.}

Particle motion drastically changes this picture. As shown in~\cite{smirnov2021}, diffusion (stochastic particle motion) can destabilize complete synchronization, leading to the establishment of a chimera state. In our previous work~\cite{Bolotov2025} for a system without elastic couplings ($\chi=0$), a regime of segregation was discovered: under the action of the force $\gamma$, a subset of particles performs transit (flying-through) motion, while the rest stays localized (trapped) in the potential wells. The phase dynamics within each of these groups self-organizes into a chimera state, forming a hierarchical chimera.
%---------------------------------------------------------------------------%
\section{Kink-driven chimera motion\label{sec:kink}}
We investigate the system of equations~\eqref{eq:main_x}--\eqref{eq:ker} with the parameter values: $L = 1$, $N=256$, $\mu=1$, $\lambda=5$, $\gamma=1.1$, $\eta=25/512 \approx 0.049$, $\psi=3$, $\kappa=5.2$, and $\alpha=1.457$. At these parameter values, Eq.~\eqref{eq:main_x} has a solutions containing two counterpropagating kinks (a kink pair), and Eq.~\eqref{eq:main_phi} supports a persisting chimera state for equidistantly fixed particle positions.

The initial conditions are set as follows. Initially, all particles are at rest and displaced by $\xi_n$ with respect to the equilibrium state where all the particles sit equidistantly at the minima $\tilde x_n$ of the substrate potential $U(x)$:
\begin{gather}
x_n(0) = \tilde x_n + \xi_n, \quad \dot x_n(0) = 0.
\label{eq:inits}
\end{gather}
To initialize a kink pair in vicinity of some potential minimum $\tilde x_d$, we set the displacement as
\begin{equation}
\xi_n = \Delta_{n - d}^{(m)}, \quad \Delta_s^{(m)} = \begin{cases}
A \cos{\left( \frac{\pi \bar s}{2m} \right)} & \text{if } \left|\bar s\right| \leq m, 
\\
0 & \text{otherwise}.
\end{cases}
\label{eq:x_pert}
\end{equation}
Here $A$ is the amplitude of exciting perturbations, $0 < m < N/2$ determines the halfwidth of the perturbation region, $\bar s = N(\{s/N +1/2\} - 1/2)$ is the ring distance from the center ($n = d$) of the perturbation region. In the calculations below, we took $A = 0.7$, $m = 3$, $d = 128$.
In all numerical runs, the initial phases $\varphi_n(0)$ are the same and correspond to an established chimera state for the unperturbed (equidistant) particle arrangement $x_n \equiv \tilde x_n$. {Note that for random initial phases $\varphi_n(0)$---for instance, those drawn from a uniform distribution on $[-\pi,\pi]$---a chimera regime is highly likely to emerge for the chosen values of parameter. This is because the motion of kinks disrupts the equidistant spacing of particles, which in turn reduces the magnitude of the mean field. This reduction increases the probability of destabilizing the fully synchronous state~\cite{smirnov2021}. However, the stability of the fully synchronous mode may be preserved with a finite probability of its occurrence.}

\begin{figure}[t!]
    \centering
    \includegraphics[width=0.9\columnwidth]{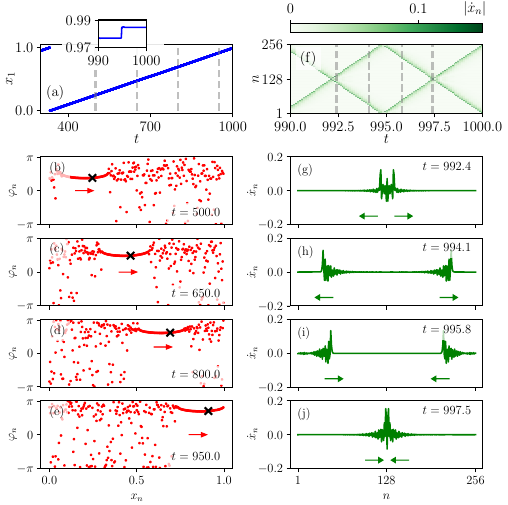}
	\caption{Results of simulations of Eqs.~\eqref{eq:main_x}--\eqref{eq:ker} with initial conditions~\eqref{eq:inits}, \eqref{eq:x_pert} corresponding to the excitation of a single kink pair driving chimera motion. Parameter values are $L = 1$, $N=256$, $\mu=1$, $\lambda=5$, $\gamma=1.1$, $\eta=25/512 \approx 0.049$, $\psi=3$, $\kappa=5.2$, $\alpha=1.457$, $A = 0.7$, $m = 3$, and $d = 128$. The calculated average velocities $\langle v_n \rangle$ of all the particles and the chimera coincide, $\langle v_{n} \rangle = \langle v_{\text{ch}} \rangle \approx 0.0013$.
    (a) Time dependence of the spatial coordinate $x_1(t)$ for an individual particle. The inset shows the time segment during which the particle transitions between adjacent potential wells of the substrate potential.
    (b--e) Snapshots of particle distribution in the coordinate--phase plane at sequential time instants: (b) $t=500$, (c) $t=650$, (d) $t=800$, and (e) $t=950$. The black crosses mark the particle with index $1$, whose spatial coordinate vs time is shown in panel (a). The red arrows indicate the direction of chimera motion (towards the average decrease of the substrate potential).
    (f) Absolute values of instantaneous particle velocities $|\dot{x}_n(t)|$ vs time for all particles. Two kinks counterpropagate and interact elastically when meeting.
    (g–-j) Snapshots of particle velocities $\dot{x}_n(t)$ at sequential time instants: (g) $t=992.4$, (h) $t=994.1$, (i) $t=995.8$, and (j) $t=997.5$. The green arrows indicate the directions of kink motion.}
	\label{fig:1}
\end{figure}

This initial state evolves into a single symmetric kink pair. The kinks in the pair counterpropagate along the chain, see Fig.~\ref{fig:1}(f)--(j). The kinks exhibit elastic interaction: upon collisions, they pass through each other while preserving their structure. During each half-period of the cyclic kink motion along the ring chain, each particle shifts to a neighboring potential minimum in the direction determined by the sign of the dc force $\gamma$, as illustrated by Fig.~\ref{fig:1}(a) and the inset there. Thus, after $N/2$ periods, each particle completes a full rotation around the system.

The phase dynamics retains its chimera character. A key observation is that the directed motion of the kinks induces a drift of the entire chimera state along the chain, as shown in Fig.~\ref{fig:1}(b)--(e). To quantify this drift, we define the instantaneous position $x_{\text{ch}}(t)$ of the chimera as the position $x_{m(t)}(t)$ of the particle experiencing the maximum (in absolute value) mean field, $\left|H_{m(t)}\right| = \max\limits_{n} \left| H_n \right|$, where
\begin{equation*}
H_n = \sum_{n'=1}^{N} G(x_{n'} - x_n) \mathrm{e}^{i\varphi_{n'}}.
\end{equation*}
In the established solution, the mean field maximum rarely switches from one particle to another, so $m(t)$ remains constant most of the time, and the chimera travels with the particles as they slowly drift along the chain. To quantify this observation, we calculated the time-average velocities $\langle v_n \rangle$ and  $\langle v_{\text{ch}} \rangle$ of particles and the chimera, defined as 
\begin{gather*}
\langle v_n \rangle = \lim_{t \to \infty} \frac{1}{t} \int_0^t 
\dot x_n \mathop{
\mathrm{d}t}, \\
\langle v_{\text{ch}} \rangle = \lim_{t \to \infty} \frac{1}{t} \int_0^t 
\dot x_{\text{ch}} \mathop{
\mathrm{d}t},
\label{eq:v_mean}
\end{gather*}
where time-derivative $x_{\text{ch}}$ is meant in terms of generalized functions since $x_{\text{ch}}$ is in general case discontinuous at the time instants where $m(t)$ changes its value. In the regime with a single kink pair, the calculated average velocities of all the particles and the chimera coincide: $\langle v_{n} \rangle = \langle v_{1} \rangle = \langle v_{\text{ch}} \rangle \approx 0.0013$.
%---------------------------------------------------------------------------%
\section{Chimera velocity quantization in the presence of multiple kink pairs\label{sec:multikink}}

We now analyze the case with the simultaneous excitation of $M > 1$ kink pairs. To this end, we use the initial conditions~\eqref{eq:inits} with
\begin{gather}
\xi_n = \sum_{j = 1}^M \Delta_{n - d_j}^{(m)}
\label{eq:multipert}
\end{gather}
with equidistantly spaced values of $d_j$---that is, we apply the initial perturbation~\eqref{eq:x_pert} $M$ times to disjoint equidistantly spaced regions of the ring chain, $(2m + 1)M < N$.

\begin{figure}[t!]
    \centering
    \includegraphics[width=1.0\columnwidth]{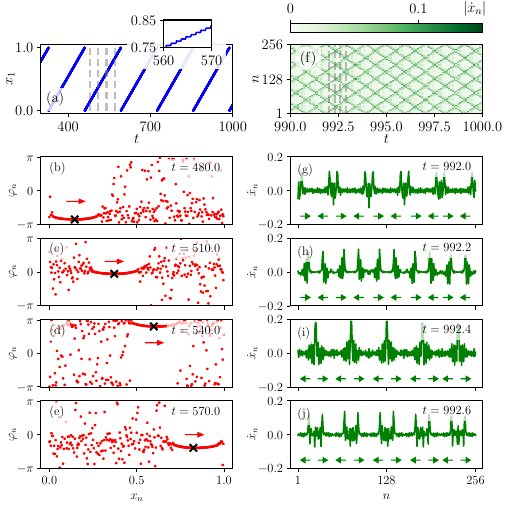}
	\caption{Same as Fig.~\ref{fig:1}, but for initial conditions~\eqref{eq:inits}, \eqref{eq:multipert} corresponding to the excitation of $M=5$ kink pairs with $d_{j} =\lfloor (2j-1)N/2M \rfloor$.  The calculated average velocities of all the particles and the chimera coincide, $\langle v_{n} \rangle = \langle v_{\text{ch}} \rangle \approx 0.0067$.} 
	\label{fig:2}
\end{figure}

\begin{figure}[t!]
    \centering
    \includegraphics[width=0.5\columnwidth]{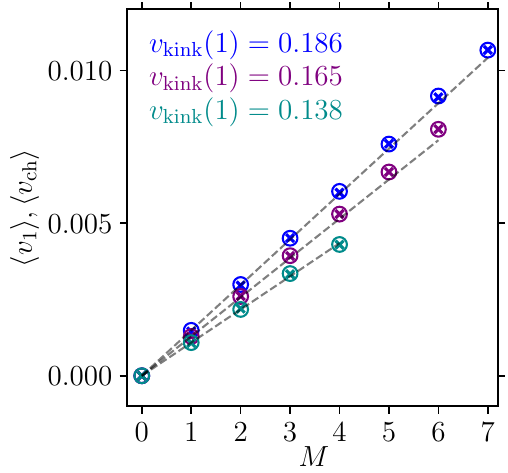}
	\caption{Calculated dependences of the average particle velocity $\langle v_1 \rangle$ (circular markers) and the average chimera velocity $\langle v_{\text{ch}} \rangle$ (cross markers) on the number $M$ of kink pairs for several values of $\psi$: $\psi = 2$ (blue markers), $\psi = 3$ (purple markers), and $\psi = 4$ (cyan markers). All other parameters (except $\psi$ and $M$) are the same as in Figs.~\ref{fig:1} and \ref{fig:2}. Different values of $\psi$ correspond to different kink velocities $v_{\mathrm{kink}}(M) = N\langle v_1\rangle/2$, which are indicated in the respective color for $M = 1$.
    The dashed straight lines represent the linear dependence $M v_k(1)$.
    On the right, each line ends at the critical value $M = M^*$---the maximum number of symmetric kink pairs which coexisted without merging for the initial conditions~\eqref{eq:inits}, \eqref{eq:multipert} used.}
	\label{fig:3}
\end{figure}

For small values of $M$, the solution of Eq.~\eqref{eq:main_x} supports stable kink dynamics: $2M$ kinks propagate and interact elastically, preserving their total number, see Fig.~\ref{fig:2}(f)--(j) corresponding to the same parameter values as used in Sec.~\ref{sec:kink} and Fig.~\ref{fig:1} but with $M = 5$. An increase in the number of kink pairs $M$ leads to a monotonic nearly proportional growth in the average particle velocity $\langle v_n \rangle$ and, consequently, in the drift velocity $\langle v_{\text{ch}} \rangle = \langle v_n \rangle \approx 0.0067$ of the chimera state; compare Figs.~\ref{fig:1}(b)--(e) and \ref{fig:2}(b)--(e). The calculated dependencies of $\langle v_n \rangle$ and $\langle v_{\text{ch}} \rangle$ on $M$ is nearly linear (Fig.~\ref{fig:3}), indicating an additive contribution of each kink pair to the overall particle transport. Nonlinear deviations from the linear trend at large $M$ are apparently caused by interactions between kinks.

Thus, the initial state of the motion subsystem governed by Eq.~\eqref{eq:main_x} determines the established velocity of the chimera drift, which can take only discrete number of values corresponding to the different topological excitation of the chain.
In this regard, the chimera velocity is quantized with elementary quantum corresponding to a kink pair.

\begin{figure}[t!]
    \centering
    \includegraphics[width=1.0\columnwidth]{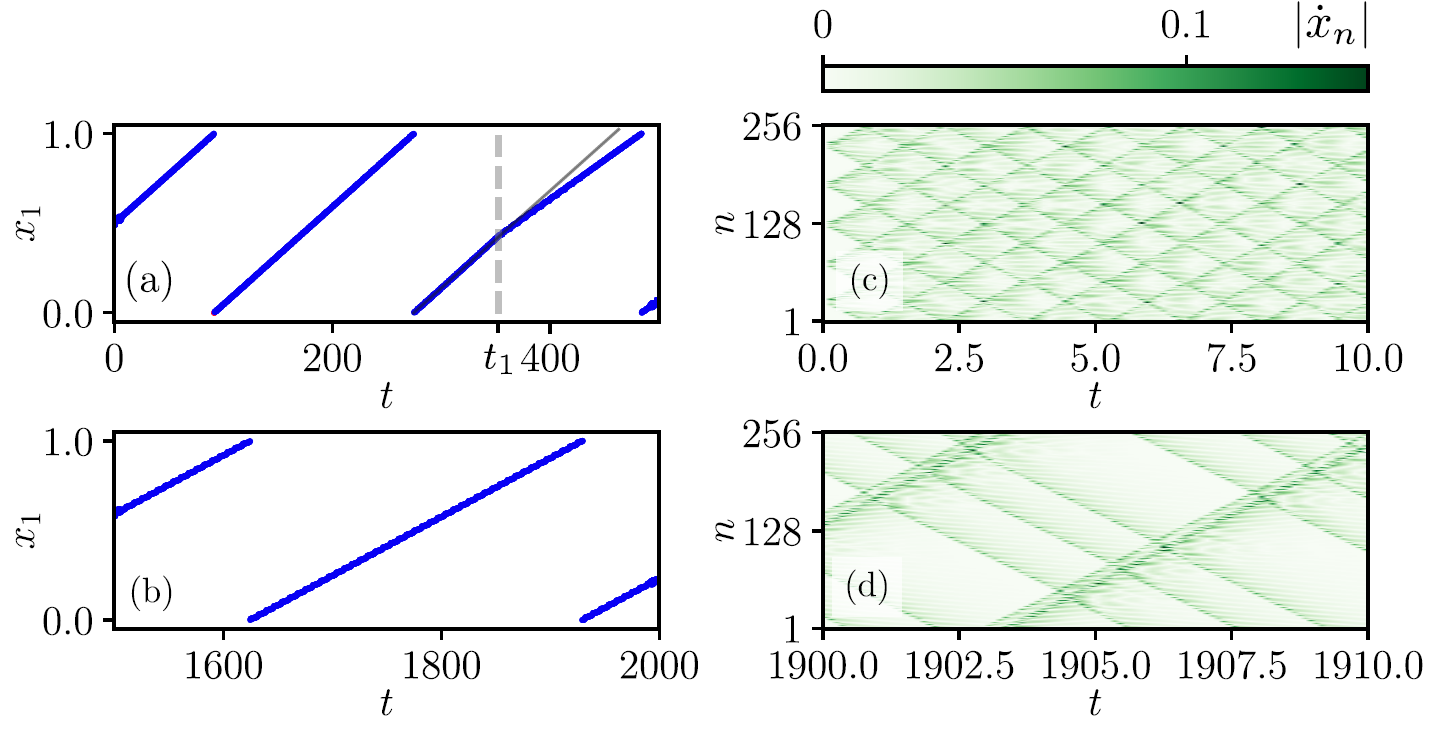}
	\caption{Results of simulations of Eq.~\eqref{eq:main_x} with initial conditions~\eqref{eq:inits}, \eqref{eq:multipert} corresponding to the initial excitation of $M = 5$ kink pairs, which evolve into 3 asymmetric kink pairs through sequential kink merging. All the parameters are the same as in Fig.~\ref{fig:2} except $\lambda=4$ and $\gamma=0.58$. (a,~b) Time dependence of the spatial coordinate $x_1(t)$ for an individual particle. At $t = t_1$, the first instance of kink destruction occurs, leading to a noticeable decrease in time-average particle velocity (i.e., a reduction in the average slope). (c,~d) Absolute values of instantaneous particle velocities $|\dot{x}_n(t)|$ vs time for all particles. Five initial symmetric kink pairs (c) finally transform into three asymmetric kink pairs (d).}
	\label{fig:4}
\end{figure}

It should be noted that beyond a critical number $M^*$ of kink pairs, their coexistence became unstable. For $M > M^*$, nonlinear kink interactions led to the annihilation of some kinks and subsequent restructuring of the regime. For example, if the previously used parameters $\lambda = 5$, $\gamma = 1.1$ are changed to $\lambda=4$, $\gamma=0.58$, the system initialized with $M=5$ no longer supports five kink pairs. Instead, it evolves into a regime with only three asymmetric kink pairs through sequential kink merging (see Fig.~\ref{fig:4}).
After each kink destruction even, the average particle velocity $\langle v_n \rangle$ drops, as seen for example at time $t=t_1$ in Fig.~\ref{fig:4}(a), and the average chimera velocity $\langle v_{\text{ch}} \rangle$ drops as well (not shown in the figure).
The resulting chimera speed is defined then by the established number of kinks and is about 3/5 of the initial value.

We should also note that, formally, Eq.~\eqref{eq:main_x} also supports regimes with collective transit motion, where the chimera velocity significantly exceeds the elementary velocity quantum, as well as nonphysical regimes in which particles with adjacent numbers cease to be nearest neighbors in physical space. Such regimes are not considered here, as they require more appropriate models---such as those incorporating nonlocal interactions or alternative particle interaction potentials like the Lennard--Jones potential.
%---------------------------------------------------------------------------%
\section{{Kink Dynamics under Random Initial Conditions}\label{sec:multikink_random}}

{As established previously, the propagation velocity of chimera patterns along the chain is determined by the speed of their carrier structures—kink-antikink pairs. Previous sections analyzed the idealized case of symmetric kinks generated by equidistant and identical initial perturbations. In this Section, we investigate a more general scenario with random initial conditions to test the robustness of the observed phenomena.
}

{
For the numerical experiment, we used model~\eqref{eq:main_x} with parameters $L = 1$, $N=256$, $\mu=1$, $\psi=3$, $\gamma = 0.58$, $\lambda = 4$, $\eta=25/512 \approx 0.049$. The initial conditions consisted of $M_0=16$ equidistantly spaced perturbations of the form~\eqref{eq:multipert} with random amplitudes $A(d_j)=A_j$ and widths $m(d_j)=m_j$.
}

\begin{figure}[t!]
    \centering
    \includegraphics[width=0.6\columnwidth]{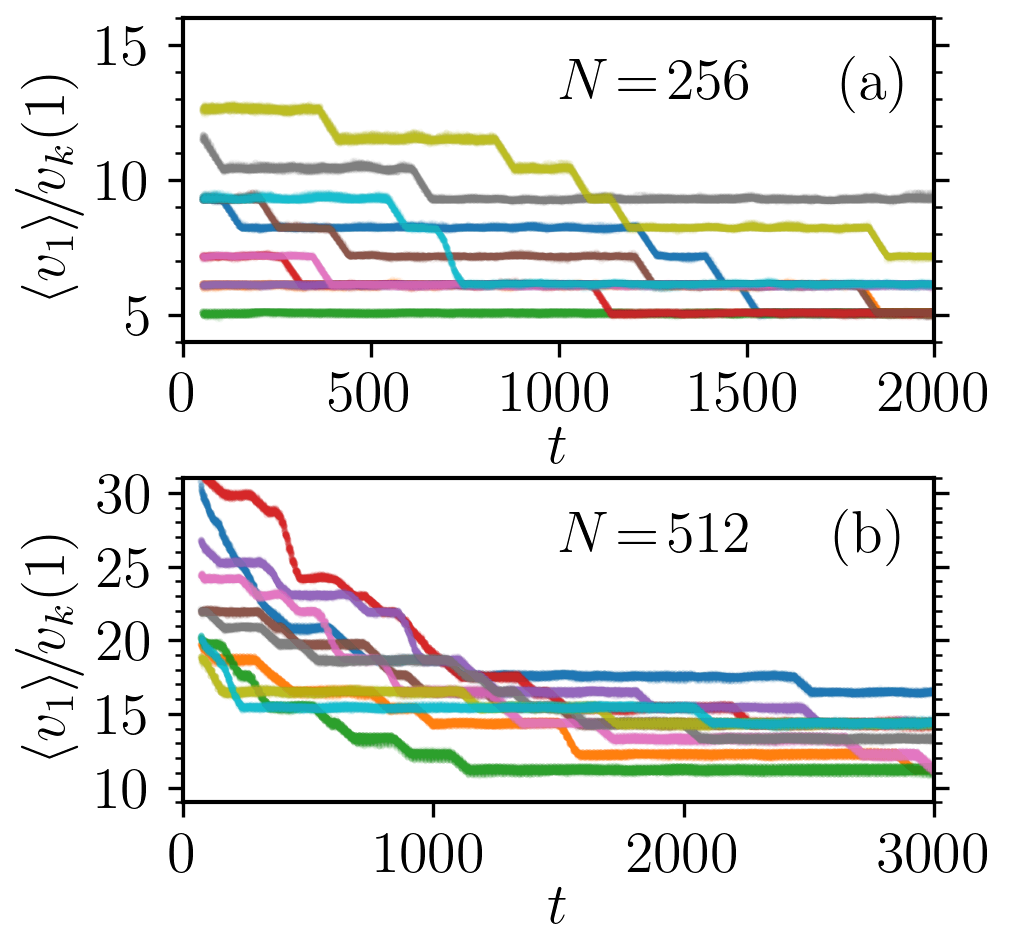}
	\caption{{
        The average particle velocity $\langle v_1 \rangle$ dynamics, normalized by the propagation velocity of a single kink pair $v_k(1)$, is plotted against time for $10$ simulations. The normalized velocity $\langle v_1 \rangle \big/ v_k(1)$ is proportional to the number of propagating kink pairs $M$. The observed decrease in $\langle v_1 \rangle \big/ v_k(1)$ indicates a reduction in the number of active kinks within the medium. The system is simulated under initial conditions of the form~\eqref{eq:multipert} with (a) $M=16$ and (b) $M=32$, equidistantly located $d_j$, random uniformly distributed $m_j \in [1, 7]$, (a) $A_j \in [0.0, 0.005]$ and (b) $A_j \in [0.0, 0.008]$.
    Parameters: $\mu=1$, $\psi=3$, $\gamma = 0.58$, $\lambda = 4$, $\eta=25/512 \approx 0.049$, and different $L$ and $N$: (a) $L = 1$, $N=256$; (b) $L = 2$, $N=512$.
    }}
	\label{fig:5}
\end{figure}

{
Figure~\ref{fig:5}a shows the temporal evolution of the mean particle velocity $\langle v_1 \rangle$, normalized by the velocity $v_k(1)$ corresponding to the propagation of a single kink pair. The system dynamics are characterized as follows: after an initial transient, several kink pairs form and may coexist. A key feature of this transient process is the annihilation of kink pairs, manifested on the figure as discrete downward jumps in the normalized velocity $\langle v_1 \rangle \big/ v_k(1)$. Each such jump corresponds to a reduction in the number of coexisting kink pairs. The system evolves toward a stationary state with a fixed number of pairs, at which point the velocity stabilizes.
}

{
A similar picture is observed in systems of larger dimension (e.g., $L=2$, $N=512$). Increasing the chain length, all other parameters being equal, leads to an increase in the average number of stable kink pairs in the stationary regime (Fig.~\ref{fig:5}b). It is important to note that for the same parameters, the final number of pairs can vary depending on the specific realization of the random initial perturbations.
}

{
For a systematic investigation, a parametric analysis was conducted for different values of dissipation and external torque. For random initial perturbations, three qualitatively distinct classes of dynamics were identified. 1) All initial perturbations decay, and the system's particles become trapped in the nearest potential wells. No kink propagation is observed. 2) A subset of the initial perturbations develops into kinks pairs. Their subsequent dynamics may include annihilations, after which the system reaches a stationary state with one or several coexisting kinks. 3) The initial perturbations lead to a state where particles undergo directed motion along the entire chain length without becoming trapped in potential wells.
}

\begin{figure}[t!]
    \centering
    \includegraphics[width=1.0\columnwidth]{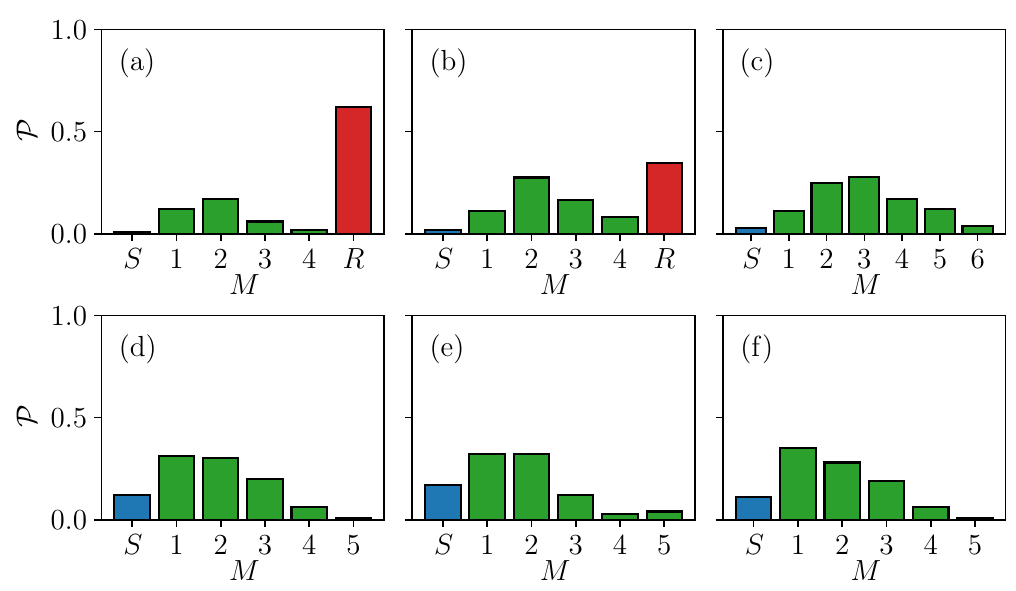}
	\caption{{
    %Probabilities $\mathcal{P}$ of realization in the Eq.~\eqref{eq:main_x} of stationary oscillators ($S$), $M$ pairs of long-lived kinks or rotational dynamics of particles ($R$) under initial conditions of the form~\eqref{eq:multipert} with $M=16$, equidistantly located $d_j$ and random uniformly distributed values of $m$ and $A$ on a segment $[1, 7]$ and $[0.0, 0.005]$. Histograms are based on $100$ simulation results. Parameters: $L = 1$, $N=256$, $\mu=1$, $\psi=3$ and and different values of $\gamma$ and $\lambda$: (a) $\gamma = 1.1$, $\lambda = 2$, (b) $\gamma = 1.1$, $\lambda = 5$, (c) $\gamma = 1.1$, $\lambda = 8$, (d) $\gamma = 0.58$, $\lambda = 2$, (e) $\gamma = 0.58$, $\lambda = 5$, (f) $\gamma = 0.58$, $\lambda = 8$.
    Probabilities $\mathcal{P}$ of the realization of stationary states ($S$, blue bins), $M=1, 2, \dots, 6$ pairs of long-lived kinks (green bins), or rotational dynamics ($R$, red bins) for Eq.~\eqref{eq:main_x}. The system is simulated under initial conditions of the form~\eqref{eq:multipert} with $M=16$, equidistantly located $d_j$, and random uniformly distributed $m_j \in [1, 7]$ and $A_j \in [0.0, 0.005]$. Histograms are built from $100$ simulation runs. Parameters: $L = 1$, $N=256$, $\mu=1$, $\psi=3$, $\eta=25/512 \approx 0.049$, and different $\gamma$ and $\lambda$: (a) $\gamma = 1.1$, $\lambda = 2$; (b) $\gamma = 1.1$, $\lambda = 5$; (c) $\gamma = 1.1$, $\lambda = 8$; (d) $\gamma = 0.58$, $\lambda = 2$; (e) $\gamma = 0.58$, $\lambda = 5$; (f) $\gamma = 0.58$, $\lambda = 8$.
    }}
	\label{fig:6}
\end{figure}

{
Diagrams in the parameter space (dissipation vs. external torque) illustrating the realized regimes are presented in Fig.~\ref{fig:6}. These diagrams clearly demonstrate the regions of stability for each of the described dynamic classes.
}
%---------------------------------------------------------------------------%

%---------------------------------------------------------------------------%
\clearpage
\section{Conclusions}\label{sec:Conclusion}
In this study, we identify and demonstrate a new scenario for organizing directed motion of a spatiotemporal structure in a system of interacting active particles with internal degrees of freedom in a spatially periodic potential. This is exemplified by a chimera state in the Kuramoto--Battogtokh system with particle positions governed by the Frenkel--Kontorova model. A localized perturbation of the equilibrium particle positions can lead to the emergence of a kink pair. The propagation of these kinks induces directed particle motion with a constant average velocity. This particle drift carries the chimera state with it, and the average chimera velocity coincides with the time-average particle velocity. Propagation of several kink pairs causes the particles to move with an average velocity that is nearly proportional to the number of these pairs. Consequently, the average chimera velocity turns out to be quantized and can be controlled by the number of localized perturbations in the initial particle arrangement---that is, by the initial conditions. The chimera velocity in this regime is orders of magnitude lower than the velocity of the transit particle motion above the potential.
The possible number kink pairs in a stable configuration is limited. Beyond this limit, sequential destruction of kink pairs occurs, accompanied by a stepwise decrease in the chimera velocity.
During this process, an initially symmetric kink configuration may transform into an asymmetric one with a reduced number of kinks.
The results obtained for the quadratic interaction potential indicate that similar dynamic scenarios might be realized in systems with physically relevant potentials, such as the Lennard--Jones potential, and in other regular systems of interacting particles with internal degrees of freedom.
%---------------------------------------------------------------------------%

\section*{Declaration of competing interest}
The authors declare that they have no known competing financial interests or personal relationships that could have appeared to influence the work reported in this paper.

\section*{Data availability}
No data was used for the research described in the article.

\section*{Acknowledgments}
This work was supported by the Ministry of Science and Higher Education of the Russian Federation under project No.~FSWR-2020-0036 (model formulation and simulations of the kink-driven chimera motion with two kinks, Secs.~\ref{sec:model} and \ref{sec:kink}) and the Russian Science Foundation under project No.~23-12-00180 (studies of the dependence on the number of kinks, Secs.~\ref{sec:multikink} and \ref{sec:multikink_random}).

 \bibliographystyle{elsarticle-num} 
 \bibliography{cas-refs}





% \end{thebibliography}
\end{document}